\documentclass[useAMS,usegraphicx,usenatbib]{mn2e}          

\usepackage{times}
\usepackage{flushend}
\usepackage{amsmath,amssymb}

\newcommand{\Msun}{\ensuremath{\,{\rm M}_\odot}}                  
\newcommand{\Rsun}{\ensuremath{\,{\rm R}_\odot}}                  
\newcommand{\Teff}{\ensuremath{T_{\rm eff}}}                      
\newcommand{\Mjup}{\ensuremath{\,{\rm M}_{\rm Jup}}}              
\newcommand{\Rjup}{\ensuremath{\,{\rm R}_{\rm Jup}}}              
\newcommand{\kms}{\,km\,s$^{-1}$}                                 
\newcommand{\as}{\ensuremath{^{\prime\prime}}}                    
\newcommand{\FeH}{\ensuremath{\left[\frac{\rm Fe}{\rm H}\right]}} 

\newcommand{\gaia}{\textit{Gaia}}
\newcommand{\reff}[1]{{#1}}
\newcommand{\refff}[1]{{#1}}

\title[A white dwarf in planetary system WASP-98]
      {A white dwarf bound to the transiting planetary system WASP-98\footnote{Based on observations made with the Southern African Large Telescope (SALT).}}

\author[Southworth et al.]
       {John Southworth\,$^{1}$, Pier-Emmanuel Tremblay\,$^{2,3}$, Boris T.\ G\"ansicke\,$^{2,3}$, Daniel Evans\,$^{1}$, \newauthor Teo Mo\v{c}nik\,$^{4}$ \\
        $^1$\,Astrophysics Group, Keele University, Staffordshire, ST5 5BG, UK \\
        $^2$\,Department of Physics, University of Warwick, Coventry CV4 7AL, UK \\
        $^3$\,Centre for Exoplanets and Habitability, University of Warwick, Coventry CV4 7AL, UK \\
        $^4$\,Gemini Observatory Northern Operations, 670 N.\ A'ohoku Place, Hilo, HI 96720, USA
        }

\begin{document} \maketitle 


\begin{abstract}
WASP-98 is a planetary system containing a hot Jupiter transiting a late-G dwarf. A fainter star 12\as\ distant has previously been identified as a white dwarf, with a distance and proper motion consistent with a physical association with the planetary system. We present spectroscopy of the white dwarf, with the aim of determining its mass, radius and temperature and hence the age of the system. However, the spectra show the featureless continuum and lack of spectral lines characteristic of the DC class of white dwarfs. We therefore fitted theoretical white dwarf spectra to the $ugriz$ apparent magnitudes and \gaia\ DR2 parallax of this object in order to determine its physical properties and the age of the system. We find that the system is old, with a lower limit of 3.6\,Gyr, but theoretical uncertainties preclude a precise determination of its age. \reff{Its kinematics are consistent with membership of the thick disc, but do not allow us to rule out the thin-disc alternative.} The old age and low metallicity of the system suggest it is subject to an age-metallicity relation, but analysis of the most metal-rich and metal-poor transiting planetary systems yields only insubstantial evidence of this. We conclude that the study of bound white dwarfs can yield independent ages to planetary systems, but such analysis may be better-suited to DA and DB rather than DC white dwarfs.
\end{abstract}

\begin{keywords}
stars: planetary systems --- stars: white dwarfs --- stars: fundamental parameters --- stars: individual: WASP-98
\end{keywords}


\section{Introduction}                                                                                                              \label{sec:intro}

The multiplicity of stars hosting transiting planets is an important piece in the jigsaw of our understanding of these systems. The majority of solar-type stars are members of binary systems \citep{DuquennoyMayor91aa,Raghavan+10apjs,DucheneKraus13araa} and this is likely to have a significant effect on the incidence and properties of planets orbiting solar-type stars. Several studies have suggested that the formation of planets is affected by the binarity of the host star \refff{\citep{DesideraBarbieri07aa}}: either hindered \citep{Fragner++11aa,Roell+12aa,Wang+15apj,Ziegler+19aj} or helped \citep{Zhang+18apj,Ziegler+19aj} depending on the orbital separation of the two stars. Some observational studies have found a lower binary fraction for stars hosting planets versus those without planets \citep{Roell+12aa,Kraus+16aj}, whereas others have found an enhanced binary fraction \citep{Ngo+16apj}. \reff{\citet{MoeKratter19xxx} found that planet formation was unaffected by companions more distant than approximately 200\,au.} Once formed, the presence of the second star is expected to affect the orbital migration of the planet \citep{WuMurray03apj,FabryckyTremaine07apj} in a way that might be observationally testable \citep{MortonJohnson11apj2}. \reff{These studies typically refer to hot Jupiters, which almost certainly form at relatively large distances from the host star and then migrate in \citep[e.g.][]{IdaLin04apj}; the prevalence of lower-mass planets may be affected by binarity in different ways.}

Of lesser scientific interest, but vital for correct determination of the physical properties of transiting extrasolar planets, is the effect of light from a nearby star on measurements of the physical properties of the system. This applies both if the nearby star is bound or merely an asterism, and the contamination can bias observationally-derived quantities \citep[e.g.][]{Daemgen+09aa,Buchhave+11apjs}. Contaminating light acts to decrease the depth of the transit in light curves, leading to an underestimate of the radius of the planet. Analogously, stationary spectral lines from a nearby star will bias the measured orbital motion of the planet host star to lower values, so the planet appears to be less massive. These effects are undetectable from a transit light curve alone \citep{Me09mn}, unless the contamination is the dominant light source in the system \citep{Bognar+15aa,Me+20aa}, but can be straightforwardly accounted for if the properties of the nearby star are known \citep[e.g.][]{Me+10mn,Buchhave+11apjs,Me+20aa}.

A particularly good example of the problems raised by the multiplicity of planet host stars is that of WASP-20, which was found to be two stars separated by 0.26$^{\prime\prime}$ when the system was imaged at high resolution \citep{Evans++16apj}. It remains unclear which star actually hosts the planet; its measured mass and radius differ by factors of 1.3 and 3.4, respectively, between the two scenarios. In both cases the planet's measured mass and radius are significantly higher than those found without accounting for binarity \citep{Anderson+15aa}. It is important to remember that binarity not only complicates the characterisation of planets, but also decreases the probability of the planet being detected in the first place, due to the dilution of the photometric and spectroscopic signals of its existence.

\begin{figure*} \includegraphics[width=\textwidth,angle=0]{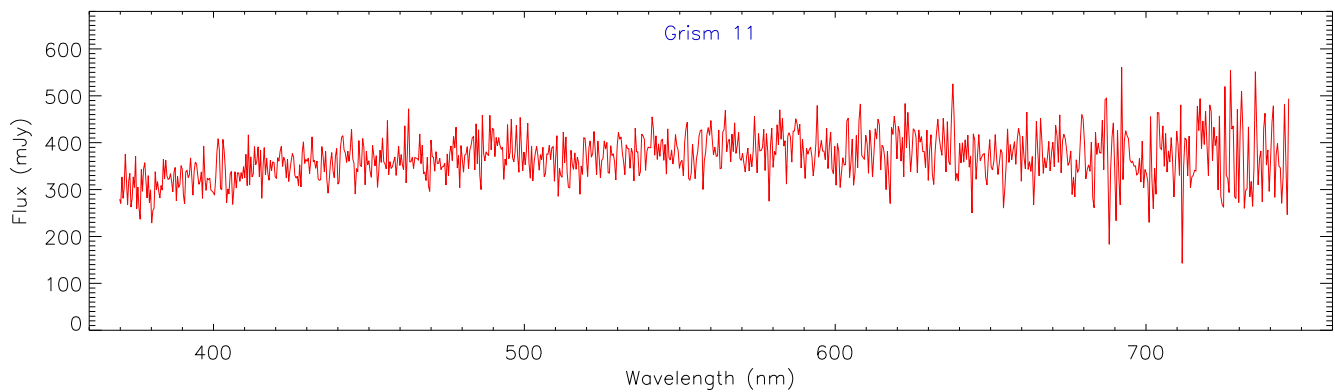}
\caption{\label{fig:spec:grism11} NTT/EFOSC2 spectrum of WASP-98\,B obtained
using grism 11. The spectrum shown is the mean of three individual spectra
after flux-calibration using the standard star LTT\,1788.} \end{figure*}

A less-travelled avenue of research is the possibility of degenerate companions to planet host stars. A consistent result from stellar evolutionary theory is that more massive stars evolve more quickly, so a wide binary system with a planet transiting the less massive star will in many cases become a transiting planetary system accompanied by a distant white-dwarf companion. \refff{The physical effects on the planet and its host star will be modest for wide binaries, for two reasons. Firstly, common-envelope evolution triggered by the evolution of the originally more massive star will preferentially remove tight ($\la$10\,au; \citealt{WillemsKolb04aa}) binary systems from the sample \citep{Paczynski71araa}. Secondly, there is an observed deficiency of planets around stars in binaries with separations of less than 200\,au (see above). }

\refff{The minimum binary separation for interactions to be negligible is not precisely known but is likely somewhere in the region of 100\,au. One example is the GJ\,86 system, which consists of a K1.5\,V star and a white dwarf with an angular separation of 1.72$^{\prime\prime}$ and an orbital separation of at least 28\,au \citep{Farihi+13mn} for which it has been postulated that the formation of the white dwarf triggered chromospheric activity in the K-star \citep{Fuhrmann+14apj}. A similar example is the K2\,V star HD\,8049, which has a white dwarf at an angular separation of 1.57$^{\prime\prime}$ (projected separation 50\,au), and a chromospheric activity and rotation speed larger than expected for its age \citep{Zurlo+13aa}. \citet{JeffriesStevens96mn} found that wind accretion from an AGB star could spin up a low-mass companion for final orbital separations of up to 100\,au. }

The effect of the white dwarf on the characterisation of the system will also be minimal: the degenerate object will be much fainter than the planet host star so the transit depth will be negligibly affected, and the faintness and spectral characteristics of the white dwarf mean it will not have a significant effect on the radial velocities measured for the host star. However, the existence of the white dwarf will have the advantage that the age of the system can be constrained from the evolution of the progenitor and the cooling time of the white dwarf \citep[see e.g.][]{Fouesneau+19apj}. So the presence of the white dwarf causes no problems in characterisation of the system but could aid in constraining its age, something that can be difficult for planet host stars less massive than the Sun due to their long evolutionary timescales, our imperfect understanding of their structure \citep{Maxted++15aa,Maxted++15aa2}, \reff{and the possibility of star-planet interactions affecting age indicators of the host star \citep{PoppenhaegerWolk14aa,Maggio+15apj}.}

In this work we present the characterisation of a white dwarf companion to the transiting planetary system WASP-98. We discuss this object and the WASP-98 system, present the observations and analysis, \reff{discuss the kinematics of the system}, and then conclude.

\begin{table} \centering \caption{\label{tab:obj} Basic information and references for the
planet host star (WASP-98\,A) and faint co-moving companion (WASP-98\,B). Masses have two
uncertainties: the first error bar is the statistical error and the second is the systematic
error from stellar models (unknown H/He abundance in the case of the white dwarf).
\newline \textbf{References:}
(1) \gaia\ DR2, pmRA and pmDec are the proper motions in RA and Dec, respectively;
(2) APASS9 \citep{Henden+12javso};
(3) \gaia\ DR2 Documentation Release 1.1, section 5.3.7;
(4) \citet{Cutri+03book};
(5) \citet{Shanks+15mn} including the nightly zeropoint accuracy;
(6) \citet{Mancini+16mn2};
(7) This work.
}
\setlength{\tabcolsep}{1pt}
\begin{tabular}{lccr} \hline
Quantity               & WASP-98\,A                  & WASP-98\,B               & Ref. \\
\hline
\gaia\ DR2 ID          & 4859136199796131200         & 4859136195500112256      & 1   \\
RA (J2000)             & 03 53 42.9627               & 03 53 42.2632            & 1   \\
Dec (J2000)            & $-$34 19 41.778             & $-$34 19 50.415          & 1   \\
$G$ magnitude          & $12.8678 \pm 0.0002$        & $20.1575 \pm 0.0062$     & 1   \\
$G_{\rm BP}$ magnitude & $13.2447 \pm 0.0013$        & $19.940  \pm 0.087$      & 1   \\
$G_{\rm RP}$ magnitude & $12.3370 \pm 0.0011$        & $19.779  \pm 0.126$      & 1   \\
Parallax (mas)         & $3.520 \pm 0.020$           &  $3.417 \pm 0.444$       & 1   \\
pmRA (mas yr$^{-1}$)   & $33.627 \pm 0.030$          & $33.381 \pm 0.738$       & 1   \\
pmDec (mas yr$^{-1}$)  & $-12.405 \pm 0.038$~~       & $-12.540 \pm 0.922$~~    & 1   \\
$V$ magnitude          & $13.063 \pm 0.026$          & 20.18                    & 2,3 \\
$J$ magnitude          & $11.691 \pm 0.027$          & --                       & 4   \\
$H$ magnitude          & $11.295 \pm 0.021$          & --                       & 4   \\
$K_s$ magnitude        & $11.284 \pm 0.023$          & --                       & 4   \\
ATLAS $u$              & $14.360 \pm 0.035$          & $20.051 \pm 0.050$       & 5   \\
ATLAS $g$              & $13.246 \pm 0.013$          & $20.089 \pm 0.015$       & 5   \\
ATLAS $r$              & $12.807 \pm 0.013$          & $20.067 \pm 0.023$       & 5   \\
ATLAS $i$              & $12.641 \pm 0.012$          & $20.174 \pm 0.036$       & 5   \\
ATLAS $z$              & $12.539 \pm 0.055$          & $20.233 \pm 0.149$       & 5   \\
\Teff\ (K)             & $5473 \pm 121$              & $8270 \pm 210$           & 6,7 \\
Mass (M$_{\odot}$)     & $0.809 \pm 0.053 \pm 0.036$ & $0.49 \pm 0.02 \pm 0.05$ & 6,7 \\
\hline \end{tabular} \end{table}


\section{The WASP-98 system}

WASP-98 was found to be a G dwarf hosting a transiting hot Jupiter by the SuperWASP survey \citep{Hellier+14mn}. Although its properties are mostly typical for a transiting planetary system with a hot Jupiter, it stood out as having an unusually low metallicity of $\FeH = -0.60 \pm 0.19$. This value was then the second-lowest measured for a transiting planet host star, after $\FeH = -0.64 \pm 0.15$ for WASP-112 \citep{Anderson+14xxx}.

The WASP-98 system was studied in more detail by \citet{Mancini+16mn2}, who obtained new high-resolution \'echelle spectroscopy plus high-precision photometry of two transits in four optical passbands. The spectroscopic analysis yielded a revised metallicity measurement of $\FeH = -0.49 \pm 0.10$, still among the lowest known for a transiting planet host star\footnote{The metallicity of WASP-98\,A measured by \citet{Mancini+16mn2} is the ninth-lowest in the TEPCat catalogue of transiting planetary systems (\citealt{Me11mn}). The catalogue was accessed on 2020/03/03 at URL \ \texttt{http://www.astro.keele.ac.uk/jkt/tepcat/}}. For reference, the physical properties returned by the analysis of \citet{Mancini+16mn2} are a mass and radius of 0.81\Msun\ and 0.74\Rsun\ for the star, and 0.92\Mjup\ and 1.14\Rjup\ for the planet.


\citet{Evans+16aa} obtained high-resolution imaging of the sky area surrounding WASP-98 as part of the HITEP lucky-imaging survey. They found one faint point source, separated from WASP-98\,A by 12$^{\prime\prime}$. It had a bluer colour than the planet host star -- blue enough to be outside the range of validity of their \Teff\ calibration -- in the two non-standard passbands of the Two Colour Imager used \citep{Skottfelt+15aa}, indicative of a \Teff\ in the region of 10\,000\,K. No further observations were obtained in the course of the project \citep{Evans+18aa}.

The publication of Data Release 2 (DR2) from the \gaia\ satellite \citep{Gaia16aa,Gaia18aa} permitted the status of nearby stars to the HITEP targets to be examined in more detail \citep{Evans18rnaas}. We found the faint nearby object (hereafter named WASP-98\,B) to have a parallax and proper motion consistent with that of the planet host star (WASP-98\,A), suggesting that the two systems were bound and thus share a common evolutionary history. Basic information on both objects is contained in Table\,\ref{tab:obj}. Based on their sky positions in \gaia\ DR2, the angular separation of the two objects is 12.23$^{\prime\prime}$ and the corresponding minimum linear separation is 3500\,au.

The properties of WASP-98\,B -- intrinsically faint yet blue in optical colour -- are consistent with it being a white dwarf. This is supported by \citet{Gentile+19mn} who constructed a catalogue of white dwarfs based on \gaia\ DR2 data. They found the object (labelled WD J035342.22$-$341950.22) to have a \Teff\ of approximately 10\,000\,K, with a probability of 0.995 of being a white dwarf. Based on this supposition we decided to obtain follow-up spectroscopy of WASP-98\,B, with the aim of determining its physical properties and thus the age of the whole system.


\begin{figure} \includegraphics[width=\columnwidth,angle=0]{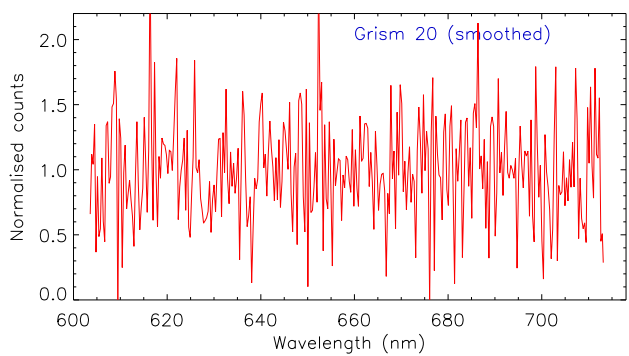}
\caption{\label{fig:spec:grism20} NTT/EFOSC2 spectrum of WASP-98\,B obtained
using grism 20. The spectrum has been rectified to a continuum level of unity
using a straight-line fit. Each group of three consecutive datapoints has been
averaged in order to reduce the noise level for visual inspection.} \end{figure}

\begin{figure*} \includegraphics[width=\textwidth,angle=0]{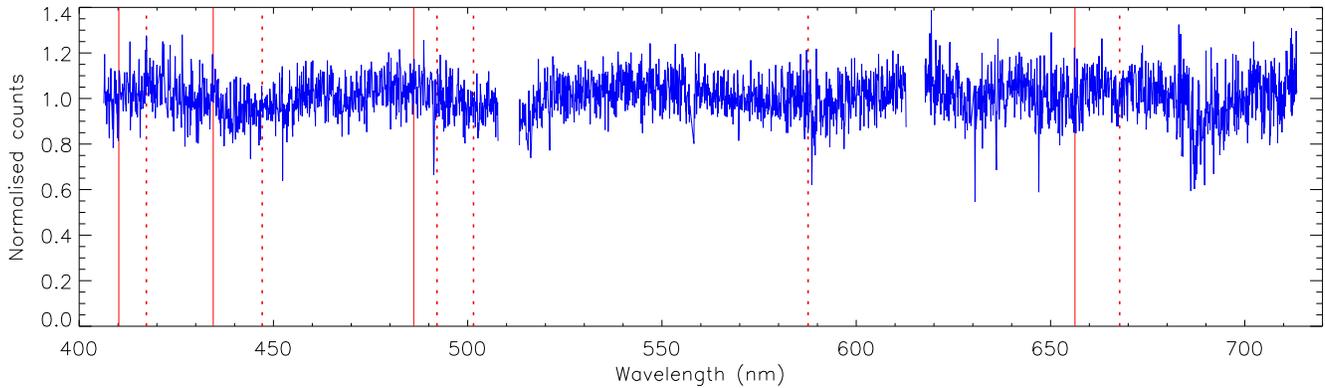}
\caption{\label{fig:spec:salt} SALT/RSS spectrum of WASP-98\,B. The spectrum has been
rectified to a continuum level of unity by dividing by a fitted quadratic polynomial.
The positions of the hydrogen Balmer lines are indicated using vertical solid lines.
A selection of helium lines typically strong in DB atmospheres \citep{Liebert77apj}
have their positions shown using vertical dotted lines.} \end{figure*}

\section{Observations}                                                                                                              \label{sec:obs}

Various surveys, most notably \gaia\ DR2 and the ESO ATLAS Public Survey, provide measured apparent magnitudes for both WASP-98\,A and WASP-98\,B (Table\,\ref{tab:obj}). WASP-98\,A is well-characterised from studies of the planet transits \citep{Hellier+14mn,Mancini+16mn2}, but little was known about WASP-98\,B. We therefore obtained follow-up spectroscopy of this object.

\subsection{NTT spectroscopy}

We used the ESO New Technology Telescope (NTT) with the EFOSC2 spectrograph \citep{Buzzoni+84msngr} to obtain spectroscopy of WASP-98\,B covering three wavelength intervals. Three consecutive spectra were taken on the night of 2018/07/09, with exposure times of 1800\,s each, using grism 11 to give a wavelength coverage of 370--746\,nm at a mean reciprocal dispersion of 0.42\,nm\,pixel$^{-1}$ (Fig.\,\ref{fig:spec:grism11}). One 1800\,s spectrum was obtained on 2018/07/08 using grism 20 to cover the H$\alpha$ line (603--713\,nm) at higher resolution (0.11\,nm\,pixel$^{-1}$) and is shown in Fig.\,\ref{fig:spec:grism20}. We also obtained a spectrum with grism 7 whilst cloud-dodging, but its S/N is too low to be useful. In all cases we used a 1$^{\prime\prime}$ slit, yielding resolutions of 1.6\,nm for grism 11 and 0.4\,nm for grism 20.

The data were reduced using a pipeline under development by the first author. The steps included bias and flat-field calibration, cosmic-ray rejection \citep{Vandokkum01pasp}, aperture extraction, wavelength calibration using a helium-argon emission lamp, and flux calibration using an observation of the standard star LTT\,1788.

The spectrum from grism 11 shows a flat continuum with no evidence for either absorption or emission lines. The spectrum from grism 20 shows no clear H$\alpha$ line absorption. This is surprising for a white dwarf, as such objects typically show strong Balmer absorption lines that would have been obvious in our spectra. The NTT spectra have a low S/N due to the faintness of the star twinned with the modest aperture of the telescope.

\subsection{SALT spectroscopy}

In an attempt to characterise WASP-98\,B via detection of weaker spectral lines that would not be apparent in the NTT spectra, we obtained a spectrum using the Southern African Large Telescope (SALT) and the Robert Stobie Spectrograph (RSS; \citealt{Burgh+03spie}) via a DDT application. The observations were obtained on the night of 2018/12/31 in service mode during dark time, and comprised two consecutive spectra each with an exposure time of 1460\,s. We used the PG900 grating with a camera tilt angle of 29.5$^\circ$, yielding a spectrum covering 406--703\,nm at a reciprocal dispersion of 0.09\,nm\,pixel$^{-1}$. The spectrum is spread over three CCD detectors so suffers from the presence of two gaps between the CCDs. The data were reduced as above, except that the wavelength calibration was obtained using an argon lamp, and no flux calibration was performed.

The two spectra were combined and data affected by strong night-sky emission lines were masked. The spectra from the individual CCDs were then each normalised to a continuum level of unity by dividing by a quadratic function. The resulting spectrum (Fig.\,\ref{fig:spec:salt}) shows no evidence for absorption lines from either hydrogen or helium, suggesting that the target star is a DC white dwarf.


\section{Analysis of the white dwarf}
\label{sec:wd}

Because the spectrum of WASP-98\,B is featureless at the S/N of our data, we have instead relied on survey photometry to determine its properties. The \gaia\ DR2 $G_{\rm BP}$  and $G_{\rm RP}$  magnitudes are unreliable due to poor background subtraction\footnote{The BP/RP excess factor (background subtraction parameter) is 1.499, which is large according to section 5.5.2 of the \gaia\ DR2 documentation: \texttt{https://gea.esac.esa.int/archive/documentation/ GDR2/pdf/GaiaDR2\_documentation\_1.2.pdf}.}. This is likely due to the presence of much brighter WASP-98\,A at 12\as, but may also be caused by WASP-98\,B being close to the faint limit of \gaia. We therefore obtained $ugriz$ magnitudes from the ATLAS survey \citep{Shanks+15mn}. The interstellar extinction between us and the WASP-98 system is very low at $0.008 \pm 0.005$\,mag\footnote{\texttt{https://stilism.obspm.fr/}} \citep{Lallement+14aa,Lallement+18aa}. We fitted the dereddened photometry but emphasise that the correction is below the statistical error bars.


\begin{figure} \includegraphics[width=\columnwidth,angle=0]{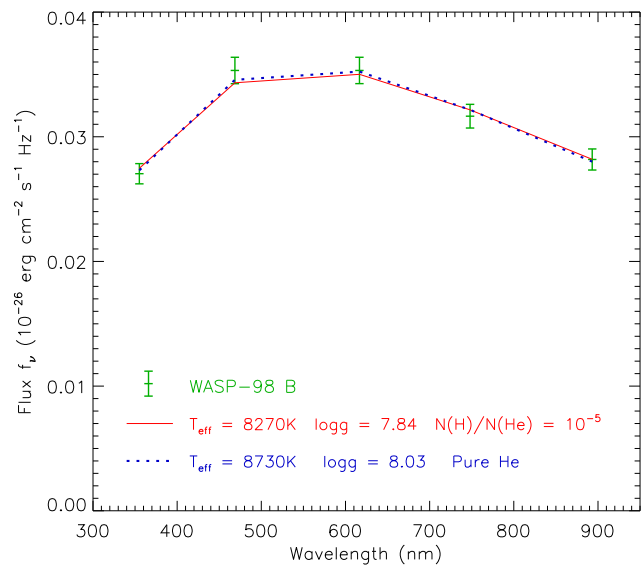}
\caption{\label{fig:WD:fit} Fit of ATLAS $ugriz$ photometry of WASP-98\,B with theoretical
white dwarf model atmospheres, one helium-only and one with H/He $= 10^{-5}$ by number.
The best-fit atmospheric parameters are identified on the panel.} \end{figure}

We converted ATLAS AB magnitudes to the equivalent SDSS system following Eq.\,(1) of \citet{Gentile+17mn} (see also \citealt{Shanks+15mn}). \citet{Gentile+17mn} found that for a sample of overlapping ATLAS and SDSS point sources with $g < 19.5$,  mean differences between corrected ATLAS and SDSS magnitudes are smaller than combined statistical uncertainties. We therefore fitted the converted ATLAS photometry with our grid of model atmosphere fluxes integrated over nominal SDSS $ugriz$ passbands.

\citet{Gentile+19mn} identified a bifurcation between hydrogen- and helium-atmosphere white dwarfs in the \gaia\ HR diagram for the range $0.0 \lesssim G_{\rm BP}\!-\!G_{\rm RP} \lesssim 0.7$ corresponding to 11\,000 K $\gtrsim \Teff \gtrsim$ 7000 K. This bifurcation is poorly reproduced by evolution tracks assuming pure-helium model atmospheres for the helium-rich population. \citet{Bergeron+19apj} have shown that by relying on helium-rich atmospheres with trace amounts of hydrogen, they could provide a much better fit to the helium-rich white dwarf cooling sequence.
From our observations of WASP-98\,B the upper limit on H/He is $\approx$ 10$^{-3.5}$ in number of atoms (see also figure 8 of \citealt{Rolland++18apj}). The immediate progenitors of DC white dwarfs are thought to be warmer DB and DBA stars with helium and hydrogen lines, for which the median H/He value is $\approx$\,10$^{-5}$ \citep{Rolland++18apj}. We therefore used this atmospheric composition as the most likely scenario for WASP-98\,B.

Fig.\,\ref{fig:WD:fit} presents our best fit to the ATLAS photometry of WASP-98\,B using mixed H/He = 10$^{-5}$ model atmospheres and the mass-radius relation of \citet{Fontaine++01pasp} for thin hydrogen layers. We obtained $\Teff = 8270 \pm 210$ K, $\log g = 7.84 \pm 0.06$ and $M = 0.49 \pm 0.02$\Msun\ with the errorbars corresponding to statistical uncertainties. Varying the fixed hydrogen abundance by $\pm$1\,dex does not significantly change our solution. However, pure-helium model atmospheres lead to a much more significant shift to $\Teff = 8730 \pm 220$\,K, $\log g = 8.03 \pm 0.07$ and $M = 0.59 \pm 0.04$\Msun.

The cooling age of the white dwarf is $0.83 \pm 0.09$~Gyr and $0.92 \pm 0.09$~Gyr with the mixed and pure-helium solutions, respectively \citep{Fontaine++01pasp}. The mass of WASP-98\,B is possibly significantly lower than the \gaia\ field white dwarf average mass of 0.58--0.59\Msun\ \citep{Tremblay+19mn}, which suggests a main-sequence progenitor with a relatively long lifetime. The mixed H/He solution for WASP-98\,B is problematic as the derived white dwarf mass is smaller than those of old halo white dwarfs \citep{Kalirai12nat}, which represent a lower limit on the mass of the oldest white dwarfs produced through single star evolution in our Galaxy. In other words, we would infer a total age well above $\approx$12\,Gyr which is unlikely given the kinematics of WASP-98\,B. It is more likely that our white dwarf mass is underestimated owing to systematic issues with the photometry or an incorrect assumption in the modelling.

We therefore included an additional systematic mass error of 0.05\Msun, to allow for a small hydrogen abundance (approximately $10^{-7}$), without including the unphysical case of a pure-helium composition as this is ruled out by \gaia\ DR2 for most DC white dwarfs. In this case the lower limit to the age of the system becomes 3.6 Gyr, using the initial-to-final mass relation of \citet{Kalirai+08apj} and the main-sequence lifetime from \citet{Hurley++00mn}. \reff{The initial mass of the white dwarf can be constrained to be between 0.80\Msun\ \citep{Kalirai12nat} and 1.53\Msun\ \citep{Kalirai+08apj} based on our lower and upper limits for the final white dwarf mass, respectively. The effect of the low metallicity of the system on its age and initial mass inferred from the initial-to-final mass relation is significantly below the uncertainties in these quantities \citep{Kalirai+09apj,Cummings+18apj}.}

The ATLAS survey has a relatively small overlap with known spectroscopic white dwarf catalogues. Only a handful of ATLAS sources have been confirmed as DC white dwarfs from spectroscopy (e.g.\ from the Montreal White Dwarf Database; \citealt{Dufour+17book}). It is also possible that the ATLAS photometry is contaminated by WASP-98\,A. For these reasons it is difficult to conclude whether WASP-98\,B is truly over-luminous, hence larger, less massive and older than field white dwarfs of similar colours and spectral type. This leaves us with a poorly constrained total age for the system, but with an indication that the WASP-98 system may be older than the average field star.


\begin{figure} \includegraphics[width=\columnwidth]{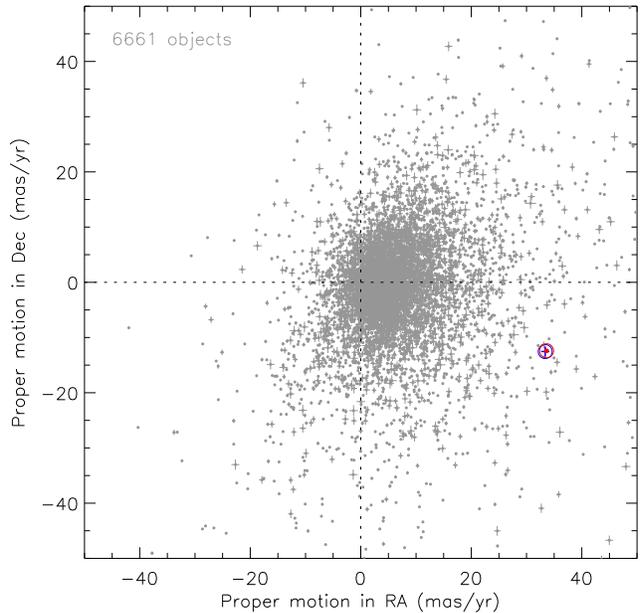}
\caption{\label{fig:pm} Plot of the proper motions in RA and Dec of all objects in
\gaia\ DR2 with measured proper motions and sky positions within 1$^\circ$ of
WASP-98\,A (grey points). WASP-98\,A and WASP-98\,B are shown using red and blue
open circles, respectively, that are almost entirely overlapping.} \end{figure}

\section{Kinematics of the WASP-98 system}

\reff{There are two points to consider for the kinematics of the system: the relation between the white dwarf and the planet host star, and the galactic population they belong to. To address the first point we obtained a catalogue of all detected sources within 1$^\circ$ of the sky position of WASP-98\,A from \gaia\ DR2, and plotted their proper motions in Fig.\,\ref{fig:pm}. The white dwarf and the planet host star are highlighted in this figure using blue and red points. It can be seen that the proper motions of the two objects are extremely similar, which supports their physical association. Their proper motions are also significantly offset from the bulk motion of the other stars in the plot, suggesting that they have a much higher space velocity than most other stars in that sky region.}

\reff{We determined the galactic space velocity of the system using the proper motion and systemic velocity of the planet host star \citep{Gaia18aa,Hellier+14mn} and the method of \citet{JohnsonSoderblom87aj} \footnote{To determine the space motions we used the {\sc gal\_uvw} procedure available in the NASA ASTROLIB library of IDL routines.}. We found $U = -17.9$\kms, $V = -5.5$\kms\ and $W = 64.4$\kms; these values have been corrected to the local standard of rest as found by \citet{Coskunoglu+11mn}. We determined the probability that the system consists of thin or thick disc stars using the method of \citet{Bensby+05aa} and the approach of \citet{Coskunoglu+11mn}, finding a 77\% probability that it is thick-disc and a 23\% probability that it is thin-disc. We are not able to definitively assign it to a single galactic population on this evidence alone.}

\reff{\citet{MackerethBovy18pasp} determined approximate values of the galactic orbital parameters of the 6.6 million stars with radial velocity measurements in \gaia\ DR2. For WASP-98 they found a small orbital eccentricity of $0.097 \pm 0.002$ and a large maximum distance from the galactic plane of $1414 \pm 54$\,pc. The uncertainties in these parameters are underestimated as they do not include the dominant contribution from the choice of galactic potential \citep{MackerethBovy18pasp}. This maximum distance is far greater than the scale height of the thin disc of our galaxy (approximately 300\,pc according to e.g.\ \citealt{GilmoreReid83mn}), which suggests a thick-disc membership for WASP-98. If so, its age is likely to be old; for example \citet{Fantin+19apj} found all thick disc stars to have formed at $9.8 \pm 1.5$\,Gyr. This supports an age of roughly 10\,Gyr for the WASP-98 system, but is unfortunately not conclusive evidence.}


\section{Summary and discussion}

We have presented the characterisation of a faint white dwarf companion, WASP-98\,B, to a transiting planet host star, WASP-98\,A. WASP-98\,B was originally discovered by our group, and was independently found to be a white dwarf by \citet{Gentile+19mn} based on its absolute magnitude and colours in the \gaia\ passbands. The distances and proper motions of the two objects in the \gaia\ DR2 catalogue agree, suggesting that they are physically related. \reff{The proper motion of the system is relatively large, and together with the systemic velocity indicates (with a probability of 77\%) that the system is part of the thick disc of the Milky Way. The sky-projected separation of the two resolved components is 12.23\as, which corresponds to a minimum physical separation of 3500\,au. The progenitor of the white dwarf had an initial mass greater than the current white dwarf mass, so the conservation of orbital energy requires the initial orbit to be smaller. The upper limit on the initial mass is 1.53\Msun\ (Section\,\ref{sec:wd}), which corresponds to a lower limit on the initial orbital separation of 1350\,au.}

Our intention was to determine the mass, radius and temperature of WASP-98\,B in order to use it as an age estimator for the planetary system. We therefore obtained spectroscopy of this white dwarf using the NTT and SALT, but found no evidence for any spectral absorption or emission lines. WASP-98\,B is therefore a DC white dwarf, whose properties had to be determined by fitting theoretical spectra to its $ugriz$ magnitudes and the distance of the system from the \gaia\ DR2 parallax of WASP-98\,A. We found a good fit to these measurements for two different sets of theoretical spectra, differing in the amount of hydrogen adopted when constructing the model atmosphere. Both alternatives indicate that the progenitor was of relatively low mass, and thus the white dwarf is old. We place a lower limit on its age of 3.6\,Gyr and caution that the upper limit is well beyond the age of the Universe.

\reff{We conclude that the WASP-98 system is old, in agreement with its likely membership of the thick-disc population, a finding that is a useful addition to our understanding of the planetary component.} \citet{Mancini+16mn2} obtained and modelled high-precision photometry of two transits in the planetary system. The low mass of the star precluded a useful constraint on its age from its density (which is measured from the transit shape), which was found to be $2.7\,^{+6.2}_{-2.5} \mbox{\,(stat.)\,} ^{+2.9}_{-1.0}\mbox{\,(sys.)\,}$\,Gyr. The current work therefore provides the first informative constraint on the age of this system.

The old age and low metallicity of the WASP-98 system is interesting in the context of the age--metallicity relation in our Galaxy expected from the chemical evolution of stellar populations \citep{FeltzingChiba13newar}. The reality of this relation is unclear, as it appears to exist in the thick-disc but not thin-disc populations \citep{Casagrande+11aa,Bensby++14aa,Casagrande+16mn}. Do the known transiting planet host stars exhibit an age-metallicity relation? To test this we queried the TEPCat catalogue \citep{Me11mn} for objects with [Fe/H] $\leqslant -0.4$ and [Fe/H] $\geqslant +0.4$, and then searched the literature for published age estimates for the systems identified. The low-metallicity sample contains 16 stars of which 15 have age estimates: the mean age is 7.7\,Gyr with a median of 10.0\,Gyr and a standard deviation of 3.7\,Gyr. The high-metallicity sample contains 15 stars, of which 14 have age estimates: the mean age is 5.8\,Gyr with a median of 7.6\,Gyr and a standard deviation of 3.6\,Gyr. The difference in ages between these two samples is not statistically significant, and the ages themselves have been estimated in a wide variety of ways (including consideration of the [Fe/H] values in some cases). A detailed analysis of a possible age--metallicity relation in planet host stars is advocated, based on homogeneous age estimates for the host stars. It would need to account for the different stellar mass distributions in the low-metallicity and high-metallicity samples, metallicity-dependent detection biases, the known bias towards higher metallicities for the host stars of giant planets \citep{Gonzalez97mn,ValentiFischer05apjs,WangFischer15aj}, \reff{and the enhancement of $\alpha$-elements in lower-metallicity stars relative to the solar abundance pattern.}

Whilst the current work was in progress, \citet{Mugrauer19mn} published a detailed analysis of 1367 extrasolar planet hosts within 500\,pc. Using \gaia\ DR2, they detected faint companions to 204 planetary systems. Within this sample are eight white dwarf companions. WASP-98\,B was included in this category based on its faintness, blue colour in optical photometry, and non-detection in 2MASS. We confirm this deduction. Finally, \citet{BonavitaDesidera20gal} recently performed a similar analysis on a sample of 850 stars with planets detected via precise radial velocity measurements. They detected ten white dwarf companions, although only nine are mentioned in their notes on individual systems, of which three were previously unknown. These companions may be useful in constraining the ages of the planetary systems, especially if they show the strong absorption lines typical of DA and DB white dwarfs.



\section*{Data availability}

The raw data underlying this article are available in the ESO Archive by searching for programme ID 0101.C-0071 or by sky position, and in the SALT Archive by searching for programme ID 2018-2-DDT-002 or by sky position. The reduced spectra will be made available at the CDS ({\tt http://cdsweb.u-strasbg.fr/}) and {\tt http://www.astro.keele.ac.uk/jkt/} and are available on reasonable request to the corresponding author.

\section*{Acknowledgements}

Based on observations made with ESO telescopes at the La Silla Paranal Observatories under programme ID 0101.C-0071 (PI: Southworth), on observations obtained with the Southern African Large Telescope (SALT) under program 2018-2-DDT-002 (PI: Southworth), and on data products from observations made with ESO Telescopes at the La Silla Paranal Observatory under program ID 177.A-3011.
We thank Luigi Mancini for helpful discussions, and an anonymous referee whose report led to a significant improvement in the quality of the manuscript.
The research leading to these results has received funding from the European Research Council under the European Union’s Horizon 2020 research and innovation programme n.677706 (WD3D).
The following internet-based resources were used in research for this paper: the ESO Digitized Sky Survey; the NASA Astrophysics Data System; the SIMBAD database operated at CDS, Strasbourg, France; and the ar$\chi$iv scientific paper preprint service operated by Cornell University.


\bibliographystyle{mn_new}

\end{document}